\documentclass[twocolumn]{aastex63}

\usepackage[caption=false]{subfig}
\usepackage{mathtools,amssymb}

\newcommand{\vsh}{v_{\rm sh}}
\newcommand{\nism}{n_{\rm ISM}}

\newcommand{\w}[1]{v_{\rm A,#1}}

\newcommand{\Emax}{E_{\rm max}}

\submitjournal{ApJ}

\shorttitle{The Maximum Energy of Shock-Accelerated Cosmic Rays}

\begin{document}

\title{The Maximum Energy of Shock-Accelerated Cosmic Rays}

\correspondingauthor{Rebecca Diesing}
\email{rrdiesing@ias.edu}

\author[0000-0002-6679-0012]{Rebecca Diesing}
\affiliation{Department of Astronomy and Astrophysics, The University of Chicago, 5640 S Ellis Ave, Chicago, IL 60637, USA}

\begin{abstract}

Identifying the accelerators of Galactic cosmic ray protons (CRs) with energies up to a few PeV ($10^{15}$ eV) remains a theoretical and observational challenge. Supernova remnants (SNRs) represent strong candidates, as they provide sufficient energetics to reproduce the CR flux observed at Earth. However, it remains unclear whether they can accelerate particles to PeV energies, particularly after the very early stages of their evolution. This uncertainty has prompted searches for other source classes and necessitates comprehensive theoretical modeling of the maximum proton energy, $\Emax$, accelerated by an arbitrary shock. While analytic estimates of $\Emax$ have been put forward in the literature, they do not fully account for the complex interplay between particle acceleration, magnetic field amplification, and shock evolution. This paper uses a multi-zone, semi-analytic model of particle acceleration based on kinetic simulations to place constraints on $\Emax$ for a wide range of astrophysical shocks. In particular, we develop relationships between $\Emax$, shock velocity, size, and ambient medium. We find that SNRs can only accelerate PeV particles under a select set of circumstances, namely, if the shock velocity exceeds $\sim 10^4$ km s$^{-1}$ and escaping particles drive magnetic field amplification. However, older, slower SNRs may still produce observational signatures of PeV particles due to populations accelerated when the shock was younger. Our results serve as a reference for modelers seeking to quickly produce a self-consistent estimate of the maximum energy accelerated by an arbitrary astrophysical shock. \footnote{Presented as a thesis to the Department of Astronomy and Astrophysics, The University of Chicago, in partial fulfillment of the requirements for the Ph.D. degree.} \\

\end{abstract}

\section{Introduction} \label{sec:intro}

After more than a century of study, the origins of the cosmic rays (CRs) detected on Earth remain uncertain. In the case of Galactic CRs, with energies up to a few PeV ($10^{15}$ eV), supernova remnants (SNRs) remain promising candidates, as they provide sufficient energetics and an efficient acceleration mechanism \citep{hillas05,berezhko+07,ptuskin+10,caprioli+10a}. 
Moreover, there is strong observational evidence that SNRs are capable particle accelerators, including the detection of hadronic $\gamma$-ray emission produced by collisions between CR protons and protons in the interstellar medium (ISM) \citep[e.g.,][]{morlino+12, slane+14, ackermann+13}.

In the standard paradigm, CRs are accelerated at the forward shocks of SNRs via \emph{diffusive shock accelration} (DSA). 
In this picture, particles scatter off of magnetic perturbations such that they diffuse back and forth across the shock, gaining energy with each crossing \citep[]{fermi54, krymskii77, axford+77p, bell78a, blandford+78}. 
This mechanism produces power-law distributions of particles and is capable of accelerating particles up to arbitrarily high energies, provided that they remain  confined close to the shock.
Moreover, DSA represents a universal mechanism that can explain particle acceleration in a multitude of astrophysical contexts \citep[e.g.,][]{ajello+21_coll, diesing+23}.

However, it is unclear whether SNRs confine CRs long enough to accelerate them up to the so-called ``knee", a feature in the CR proton spectrum around a few PeV \citep[e.g.,][]{blumer+09, argo15b, icecube13b} and the presumed maximum energy of Galactic CR protons.
Namely, in the absence of strong magnetic field amplification, SNRs are unable to confine PeV particles \citep{lagage+83b}. While there is clear theoretical \citep[e.g.,][]{bell04, amato+09, reville+12, zacharegkas+22} and observational \citep[e.g.,][]{volk+05, parizot+06, morlino+10, ressler+14, tran+15} evidence that particle acceleration can drive strong magnetic field amplification in SNRs, the question of whether that amplification is sufficient to confine PeV particles remains contested in the literature \citep[e.g.,][]{ptuskin+10,  bell+13, cardillo+15, marcowith+18, gabici+19, cristofari+20, cristofari+21, brose+22}.

Observational efforts to search for PeVatrons have also called into question the SNR paradigm for Galactic CR acceleration. 
While $\gtrsim 100$ TeV $\gamma$-ray sources have been detected with instruments such as the Large High Altitude Air Shower Observatory (LHAASO) \citep{cao+21}, the High Altitude Water Cherenkov Observatory (HAWC) \citep{HAWC20}, and the High Energy Stereoscopic System (H.E.S.S.) \citep{HESS21}, many of these sources do not appear to coincide with known SNRs. 
These results motivate the search for alternative sources of PeV protons, such as pulsar winds \citep[e.g.,][]{amato14}, microquasars \citep[e.g.,][]{abeysekara+18}, star clusters \citep[e.g.,][]{aharonian+19, bykov+20}, and superbubbles \citep[e.g.,][]{parizot+04b}.
Notably, these alternative PeVatron candidates often still involve shocks, and tend to invoke DSA as the predominant acceleration mechanism.

In this paper, we place constraints on the maximum proton energy, $\Emax$, accelerated by an arbitrary astrophysical shock using a self-consistent, multi-zone model for particle acceleration and magnetic field amplification. 
While we tailor our model parameters to SNRs in order to analyze their potential to be PeVatrons, we also develop scaling relations that can be applied to any spherical forward shock, such as novae \citep[e.g.,][]{diesing+23} and fast black-hole winds \citep[e.g.,][]{ajello+21_coll}.
We also cast these scaling relations in terms of parameters that can be inferred observationally (e.g., shock velocity).
Finally, we bracket theoretical uncertainties in the nature of magnetic field amplification, resulting in robust lower and upper limits on $\Emax$.
This combination of features---our multi-zone, self-consistent approach, the easy applicability of our results to observations, and our accounting for theoretical uncertainties---distinguishes our work from others in the literature.

This paper is organized as follows: we describe our model for shock evolution, particle acceleration, and magnetic field amplification in Section \ref{sec:method}. In Section \ref{sec:results}, we summarize the results of our $\Emax$ calculations, discussing their implications for potential PeVatron candidates and searches in Section \ref{sec:discussion}. 
We conclude in Section \ref{sec:conclusion}.

\section{Method} \label{sec:method}

To calculate $\Emax$ for an arbitrary shock, we use a multi-zone model of particle acceleration employed in \cite{diesing+21} and references therein.
Broadly speaking we choose model parameters to be consistent with typical SNRs. 
However, our results can be scaled up or down to approximate the maximum energy accelerated by other types of astrophysical shocks.
Note that, throughout this paper, subscripts 0, 1, and 2 are used to denote quantities far upstream, immediately upstream, and downstream of a shock.

\subsection{Shock Hydrodynamics} \label{subsec:hydro}

To estimate typical SNR evolution, we employ a formalism similar to that described in \cite{diesing+18}, which assumes that ejecta and swept-up material form a thin shell behind the forward shock, which expands due to pressure from a hot bubble inside it \cite[see, e.g.,][for examples of this \emph{thin-shell approximation}]{bisnovatyi-kogan+95, ostriker+88, bandiera+04}. More specifically, we model SNRs during two stages of evolution: the \emph{ejecta-dominated stage}, in which the mass of swept-up material is less than the ejecta mass and the shock expands freely, and the \emph{Sedov-Taylor stage}, in which the swept-up mass exceeds the ejecta mass and the shock expands adiabatically. 
 
Because energy conserved throughout both stages, we can write this evolution as,
\begin{equation}
    v_{\rm sh} = \bigg(\frac{2E_{\rm SN}}{M_{\rm ej}+M_{\rm SU}}\bigg)^{1/2},
\end{equation}
where, $v_{\rm sh}$ is the forward shock velocity, $E_{\rm SN}$ is the initial SN kinetic energy, $M_{\rm ej}$ is the ejecta mass, and $M_{\rm SU}$ is the swept up mass, given by,
\begin{equation}
    M_{\rm SU} = \int_{R_{\rm min}}^{R_{\rm sh}} 4\pi r^2\rho_0(r)dr.
\end{equation}

Eventually, the temperature behind the forward shock drops below $10^6$ K and the SNR becomes radiative. However, at this point, the shock has slowed substantially such that $\Emax$ has fallen well below its peak. Put another way, the DSA timescale for CRs of energy $E = \Emax$ is given by $\tau_{\rm DSA} \approx D(\Emax)/\vsh^2$ where $D(E)$ is the energy-dependent diffusion coefficient.
Assuming Bohm diffusion \citep{caprioli+14c, reville+13}, $D(E) \propto r_{\rm L} \propto E/B_2$ where $r_{\rm L} $ is the Larmor radius and $B_2$ is the postshock magnetic field. This gives $E_{\rm max} \propto B_2\vsh^2 t$ and, since $\vsh$ is roughly constant during the ejecta dominated stage, $E_{\rm max}$ initially increases, provided that $B_2$ remains constant (as is likely to be the case in a uniform ambient medium).
After the transition to the Sedov stage, the shock slows down such that $\vsh \propto t^{-3/5}$, meaning that $E_{\rm max}$ decreases with time, i.e., $E_{\rm max} \propto B_2(t)t^{-1/5}$ \citep{cardillo+15,bell+13}. This decrease in $\Emax$ happens even earlier for shocks expanding into a wind profile ($n \propto r^{-2}$), since the decrease in density also leads to a decrease in the amplified postshock magnetic field. Thus, $\Emax$ depends most strongly on shock evolution during the ejecta-dominated and early Sedov-Taylor stages, and we do not model the onset of the radiative stage, nor do we model SNRs older than $10^4$ yr.

Throughout this work, we consider a benchmark SNR with $E_{\rm SN} = 10^{51}$ erg and $M_{\rm ej} = 1 M_{\odot}$, expanding into uniform media of number density $\nism \in [10^{-2}, 10^2]$ cm$^{-3}$ (i.e., the environments around typical Type Ia SNe), and into wind profiles given by $\nism \in [10^{-1}, 10]\times 3.5(R/\text{pc})^{-2}$ cm$^{-3}$ (i.e., the environments around typical core-collapse SNe). Note that $\nism = 3.5(R/\text{pc})^{-2}$ cm$^{-3}$ corresponds to a stellar wind with velocity $10$ km s$^{-1}$ and mass-loss rate $10^{-5} M_{\odot}\text{ yr}^{-1}$ \citep[e.g., ][]{weaver+77}. To accommodate for the wide diversity of core-collapse SNe, we also consider SNRs with $E_{\rm SN} = 10^{52}$ erg and $M_{\rm ej} = 5 M_{\odot}$, all expanding into wind profiles.

\subsection{Particle Acceleration}
\label{subsec:Acceleration}

We model particle acceleration using a semi-analytic framework that self-consistently solves the steady-state diffusion-advection equation for the transport of non-thermal particles in a quasi-parallel, non-relativistic shock, including the dynamical effects of accelerated particles and CR-driven magnetic field amplification:
\begin{equation}
\begin{split}
    \tilde{u}(x)\frac{\partial f(x,p)}{\partial x } = \frac{\partial}{\partial x}\left[D(x,p)\frac{\partial f(x,p)}{\partial x}\right] \\ 
    + \frac{p}{3}\frac{d \tilde{u}(x)}{dx} \frac{\partial f(x,p)}{\partial p} + Q(x,p).
\end{split}
\label{eq:parker}
\end{equation}
Here, $\tilde{u}(x)$ is the velocity of the magnetic fluctuations responsible for scattering particles, $Q(x,p)$ accounts for particle injection, and $D(x,p) = \frac{c}{3}r_{\rm L}$ is a Bohm-like diffusion coefficient \citep[see e.g.,][]{caprioli+14c, reville+13}. For detailed descriptions of this model, see \cite{caprioli+09a,caprioli+10b, caprioli12, diesing+19, diesing+21} and references therein, in particular \cite{malkov97,malkov+00,blasi02,blasi04,amato+05, amato+06}. 

We assume that protons with momenta above $p_{\rm inj} \equiv \xi_{\rm inj}p_{\rm th}$ are injected into the acceleration process, where $p_{\rm th}$ is the thermal momentum and we choose $\xi_{\rm inj}$ to produce CR pressure fractions $\sim 10\%$, consistent with kinetic simulations of quasi-parallel shocks \citep[e.g., ][]{caprioli+14a}. We also calculate $\Emax$ self-consistently by requiring that the diffusion length (assuming Bohm diffusion) of particles with energy  $E = \Emax$ be 10$\%$ of the shock radius (see Section \ref{subsec:Emax} for a detailed discussion).

This model calculates the instantaneous spectrum of protons accelerated at each timestep of shock evolution, as well as the corresponding flux of escaping particles (see Section \ref{subsec:Emax} for a detailed discussion).
These spectra are then shifted and weighted to account for adiabatic losses \citep[see][for more details]{caprioli+10a,morlino+12,diesing+19}, before being added together to produce the cumulative, multi-zone spectrum of particles accelerated by an arbitrary shock.

It is worth noting that this model also includes the effect of the \emph{postcursor}, a drift of CRs and magnetic fluctuations with respect to the plasma behind the shock that arises in kinetic simulations \citep[][]{haggerty+20, caprioli+20}. This drift moves away from the shock with a velocity comparable to the local Alfv\'en speed in the amplified magnetic field, sufficient to produce a substantial steepening of the CR spectrum consistent with observations \citep[see][for a detailed discussion]{diesing+21}. While this effect has little bearing on the value $\Emax$, it has important consequences for the number of CRs produced at this energy. A detailed discussion of these consequences can be found in Section \ref{sec:discussion}.

\subsection{Magnetic Field Amplification
}
The propagation of CRs ahead of the shock is expected to excite streaming instabilities, \citep[]{bell78a,bell04,amato+09,bykov+13}, which drive magnetic field amplification and suppress the CR diffusion coefficient \citep{caprioli+14b,caprioli+14c}. 
The result is magnetic field perturbations with magnitudes that far exceed that of the ordered background magnetic field. 
This magnetic field amplification has been  inferred observationally from the X-ray emission of many young SNRs, which exhibit narrow X-ray rims due to synchrotron losses by relativistic electrons \citep[e.g., ][]{parizot+06, bamba+05, morlino+10, ressler+14}. 
Such magnetic field amplification is also essential for SNRs to accelerate protons to even the multi-TeV energies inferred from $\gamma$-ray observations of historical remnants \citep[e.g., ][]{morlino+12,ahnen+17}, implying that a proper treatment of magnetic field amplification is essential to predict $\Emax$.

We model magnetic field amplification by assuming pressure contributions from both the resonant streaming instability \citep[e.g., ][]{kulsrud+68,zweibel79,skilling75a, skilling75b, skilling75c, bell78a, lagage+83a}, and the non-resonant hybrid instability \citep{bell04}. 
Detailed discussions of these instabilities can also be found in \cite{cristofari+21}.

In the resonant instability, CRs excite Alfv\'en waves with a wavelength equal to their gyroradius. This instability saturates when the magnitude of the resulting magnetic perturbations reaches that of the ordered background field: $\delta B/B \sim 1$. 
\cite{amato+06} derives the magnetic pressure at saturation, $P_{\rm B1,res}$, to be,

\begin{equation}
    P_{\rm B1,res} = \frac{P_{\rm CR,1}}{4 M_{\rm A, 0}},
\end{equation}
where $P_{\rm CR,1}$ is the CR pressure in front of the shock and $M_{\rm A} \equiv \vsh/v_{\rm A,0}$ is the Alfv\'enic Mach number.

For the fast shocks considered in this work, much more significant is the non-resonant hybrid instability. 
Driven by CR currents in the upstream,  \cite{bell04} predicts that saturation occurs when the magnetic field pressure in front of the shock, $P_{\rm B1,Bell}$, reaches approximate equipartition with the anisotropic fraction of the CR pressure, yielding,

\begin{equation}
    P_{\rm B1,Bell} = \frac{\vsh}{2c}\frac{P_{\rm CR,1}}{\gamma_{\rm CR} - 1}.
\end{equation}
See also \cite{blasi+15}. Here, $c$ is the speed of light and $\gamma_{\rm CR}=4/3$ is the CR adiabatic index. This saturation occurs on timescales much shorter than those considered in this work \citep{blasi+15, zacharegkas+22}, can lead to $\delta B/B_0 \gg 1$, and dominates in SNR-like environments.

It is worth noting that some recent works find that the non-resonant instability does not saturate due to the fact that, in very young SNRs, the shock-capture time of the precursor can be short, meaning that only a few growth cycles are available to develop the non-resonant instability \citep{brose+22, inoue+21}. However, these conclusions are based on implicit assumptions about the nature of the CR current. Namely, in the case of \cite{inoue+21}, the simulation box used is not large enough to capture the current of escaping particles, leading to a truncation of the CR current near the edge of the box, and a corresponding decrease in ratio between the advection time and the growth time. In the case of \cite{brose+22}, a magnetic amplification factor is set a priori, which in turn sets the size of the shock precursor and thus the number of growth cycles available to develop the non-resonant instability. Conversely, works that treat the CR current in a more self-consistent manner \citep[e.g.,][]{bell+13, blasi+15} find that a sufficient number of growth times can indeed be achieved, justifying the assumption of saturation in this work.

That being said, a comprehensive theory for CR-driven magnetic field amplification upstream of a shock is still missing. Namely, it remains unclear whether the particles responsible for the CR currents that drive the non-resonant instability are primarily those diffusing near the shock \citep[e.g.,][]{bell04,amato+06} or those escaping in the far upstream \citep[e.g.,][]{vladimirov+06,caprioli+09b,bell+13}. On the one hand, diffusing particles with energies $E < \Emax$ are greater in number. On the other, escaping particles with energy $E = \Emax$ have a larger drift speed. Thus, the relative contribution from each population should depend on the spectral slope, as discussed in \cite{cristofari+21}. 

In order to bracket this uncertainty, we consider two extreme scenarios: 
\begin{enumerate}
    \item[A:] Diffusing CRs are solely responsible for magnetic field amplification, such that $P_{\rm B,Bell}(x) \propto P_{\rm CR}(x)$. This scenario gives a lower limit on the amplified magnetic field and thus $\Emax$.
    \item[B:] Escaping CRs are solely responsible for magnetic field amplification, such that $P_{\rm B,Bell}(x) = P_{\rm B1,Bell}$. This scenario gives an upper limit on the amplified magnetic field and thus $\Emax$.
\end{enumerate}

Assuming that all components of the magnetic perturbations upstream are compressed, the downstream magnetic field strength is given by $B_2 \simeq R_{\rm sub}B_1$, where $R_{\rm sub} = u_1/u_2$ is the subshock compression ratio. 
Our typical SNR parameters give $B_2$ near a few hundred $\mu$G, in good agreement with X-ray observations of young SNRs \citep{volk+05,parizot+06,caprioli+08}. 

Note that, throughout this work, we consider a uniform ambient magnetic field equal to that of the ISM: $B_0 = 3\mu$G. While this value may not hold in a stellar wind, changing $B_0$ has negligible impact on $\Emax$, since the ordered magnetic field is not responsible for confining particles and the turbulent field amplified via the non-resonant instability does not depend on $B_0$.

\subsection{The Maximum Energy}
\label{subsec:Emax}

As mentioned in Section \ref{subsec:Acceleration}, the maximum energy accelerated at any given time (i.e., the instantaneous $\Emax$) is set by equating the diffusion length of protons with energy $E = \Emax$ with the size of the acceleration region, taken to be  10\% the radius of the SNR, $R_{\rm sh}$, (roughly the extent of the swept-up ambient material behind the shock). 
Thus, we have $D(\Emax)/\vsh = 0.1 R_{\rm sh}$. Assuming Bohm diffusion \citep[e.g., ][]{caprioli+14c, reville+13}, we obtain, $\Emax \propto B_2\vsh R_{\rm sh}$, which is equivalent to the age-limited scaling relation derived in Section \ref{subsec:hydro}. Combined with our prescription for magnetic field amplification in the case that non-resonant instability dominates, this scaling relation becomes,
\begin{equation}
\label{eqn:Emax}
\Emax \propto \nism^{1/2}\vsh^{5/2}R_{\rm sh},
\end{equation}
assuming that the CR pressure is a fixed fraction of the ram pressure, $\propto \nism\vsh^2$. 

More specifically, when solving the transport equation for nonthermal particles, we require the distribution function to vanish at a distance $0.1 R_{\rm sh}$ upstream of the shock, mimicking the presence of a free-escape boundary beyond which particles cannot diffuse back to the shock \citep[see][for a detailed discussion]{caprioli+10b}. The instantaneous escape flux is also calculated as the flux of particles crossing this boundary.

It is worth noting that escape-limited particle acceleration may have interesting implications for the slopes of the distributions accelerated by post-adiabatic SNRs, which produce steep, broken power laws at late times \citep[e.g.,][]{ohira+10, celli+19, brose+20}. In particular, \cite{brose+20} finds that escape from deep downstream may be able to reproduce the spectra observed for relatively old SNRs such as W44 and IC443. However, at such late times ($t > 10^4$ yr), SNR shocks are likely to be radiative and/or weak, with $\Emax$ having fallen well below its peak as discussed in Section \ref{subsec:hydro}. We therefore limit our model to the ejecta-dominated and Sedov-Taylor phases of shock evolution ($t \leq 10^4$ yr), allowing us to safely neglect downstream escape and other late-time effects \citep[e.g., reacceleration; see][]{cardillo+16}.

After adding together the weighted instantaneous proton spectra as described in Section \ref{subsec:Acceleration}, we approximate the cumulative $\Emax$ as the energy associated with the maximum of the cumulative escape flux. This energy is roughly equal to the energy at which the proton distribution drops by one e-fold \citep[][]{caprioli+09b}.

\section{Results} \label{sec:results}

\begin{figure*}[ht]
    \centering
    \includegraphics[width = \textwidth, clip=true, trim={0, 15, 0, 0}]{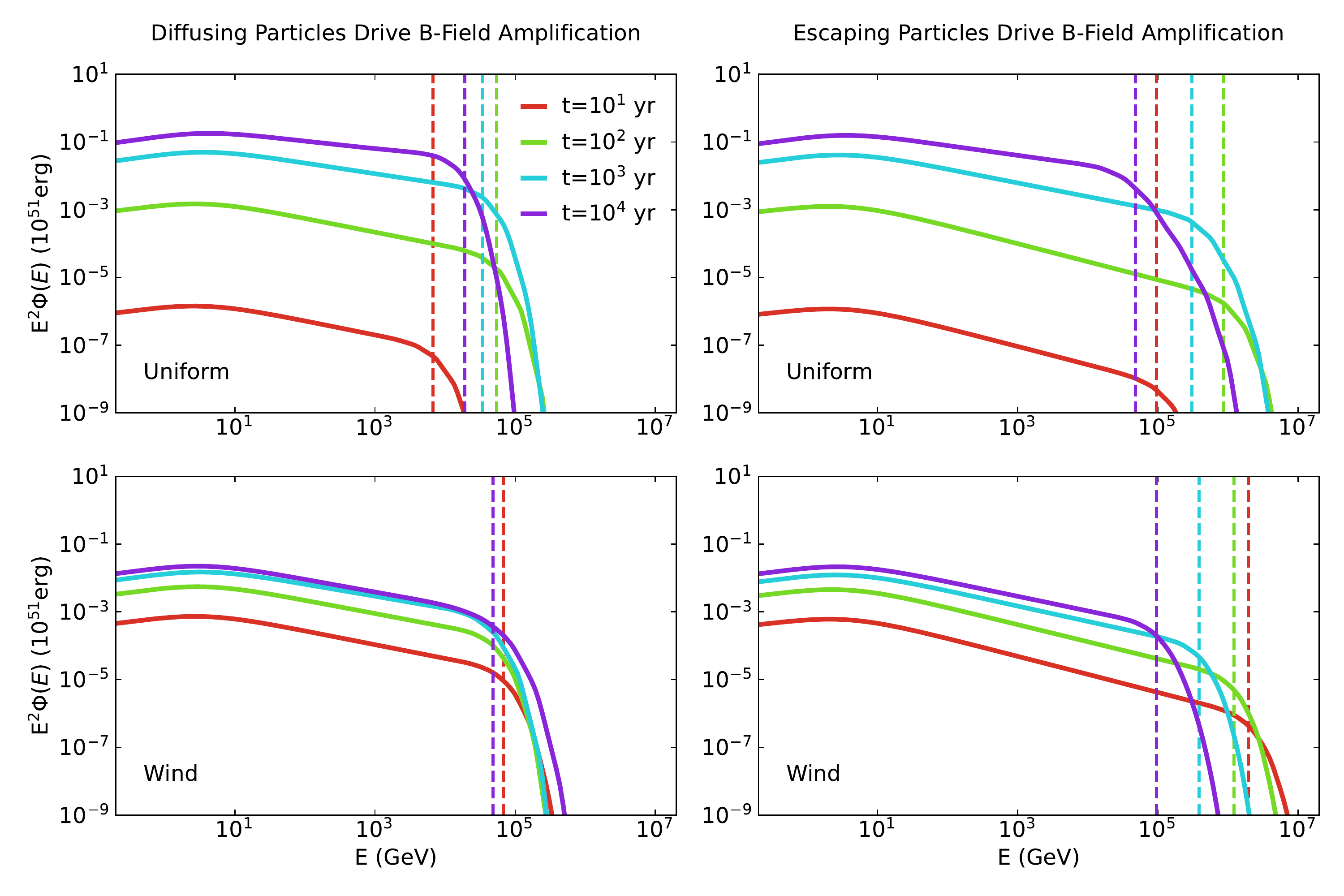}
    \caption{Cumulative spectra of accelerated protons produced by our benchmark SNR ($E_{\rm SN} = 10^{51}$ erg, $M_{\rm ej} = 1 M_{\odot}$) at different stages (denoted by line color) of shock evolution, assuming diffusing particles drive magnetic field amplification (left column) or escaping particles drive magnetic field amplification (right column). The top row of spectra corresponds to a uniform ambient medium of density 1 cm$^{-3}$, while the bottom row corresponds to a wind profile of density $3.5(R/\text{pc})^{-2}$ cm$^{-3}$. The maximum proton energy of each spectrum is denoted with a dashed vertical line. Typical SNRs can only accelerate PeV particles if escaping CRs drive magnetic field amplification, and even then only at relatively early times ($t \sim 100$ yr). \\}
    \label{fig:Spectra}
\end{figure*}

Cumulative proton spectra from our benchmark SNR are shown in Figure \ref{fig:Spectra} with the left (right) column corresponding to the case in which diffusing (escaping) particles drive magnetic field amplification. The top and bottom rows correspond to expansion into a typical uniform medium ($\nism = 1\text{ cm}^{-3}$) and wind profile ($\nism = 3.5(R/\text{pc})^{-2}\text{ cm}^{-3}$), respectively. In these cases, $\Emax$ reaches PeV energies only under select conditions. Namely, SNRs can only be PeVatrons under the optimistic assumption that escaping particles drive magnetic field amplification and, even then, can only produce PeV particles for a brief period at roughly $t \sim 100$ yr for the uniform case and prior to $t \sim 100$ yr for the wind case. This result is broadly consistent with results in the literature \citep[e.g., ][]{bell+13, cristofari+21}, which find that typical SNRs are not likely to be PeVatrons.

\begin{figure*}[ht]
    \centering
    \includegraphics[width = \textwidth, clip=true, trim={0, 15, 0, 0}]{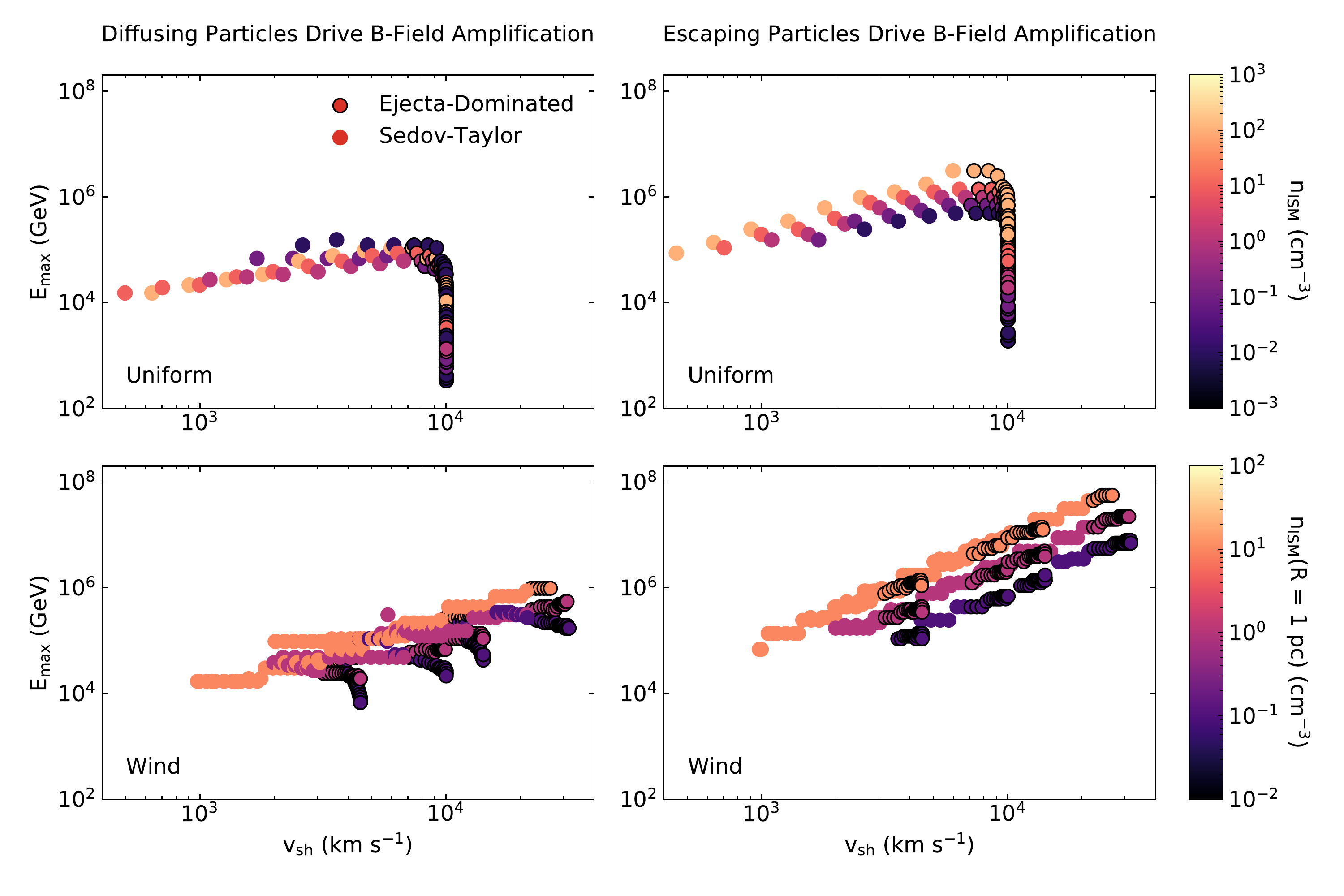}
    \caption{$\Emax$ as a function of shock velocity, ($\vsh$), for a variety of SNRs expanding into media of different ambient density normalizations (color scales), assuming diffusing particles drive magnetic field amplification (left column) or escaping particles drive magnetic field amplification (right column). The top row corresponds to expansion into uniform ambient media and, to be broadly consistent with an SNR from a Type Ia SN, only considers our benchmark scenario. Meanwhile, to capture the wider range of parameters associated with core-collapse SNe, the bottom row corresponds to expansion of SNRs into wind profiles with $E_{\rm SN} \in [1,10] \times 10^{51}$ erg and $ M_{\rm ej} \in [1, 5] \ M_{\odot}$. Outlined points denote SNRs that are still ejecta-dominated. For SNRs expanding into wind profiles, evolutionary stage has little bearing on $\Emax$ such that $\vsh$ serves as a good predictor of its value (modulo density normalization). On the other hand, for SNRs expanding into uniform media, $\vsh$ is only a good predictor of $\Emax$ during the Sedov-Taylor phase. \\}
    \label{fig:EmaxVsh}
\end{figure*}

We also note that, in these benchmark cases, we achieve maximum energies that are higher than those observationally inferred for some historical SNRs, in particular Cas A \citep{abeysekara+20}. This discrepancy likely arises from the complex evolutionary history of some core-collapse SNe \citep[see, e.g.,][in the case of Cas A]{orlando+22}, and indicates that our benchmark cases are not representative of all SNRs. An estimate of $\Emax$ that includes a broader range of evolutionary scenarios can be found in Figure \ref{fig:EmaxVsh}.

Note that the spectra shown in Figure \ref{fig:Spectra} are steeper than the canonical $\Phi(E) \propto E^{-2}$ predicted by standard DSA \citep[e.g.,][]{bell78a}. This steepening arises from the \emph{postcursor} effect described in Section \ref{subsec:Acceleration}, and is consistent with $\gamma$-ray observations of historical SNRs \citep[e.g.,][]{giordano+12,archambault+17,saha+14}. 

\begin{figure}[ht]
    \centering
    \includegraphics[width = 0.47\textwidth, clip=true, trim = {15, 15, 10, 10}]{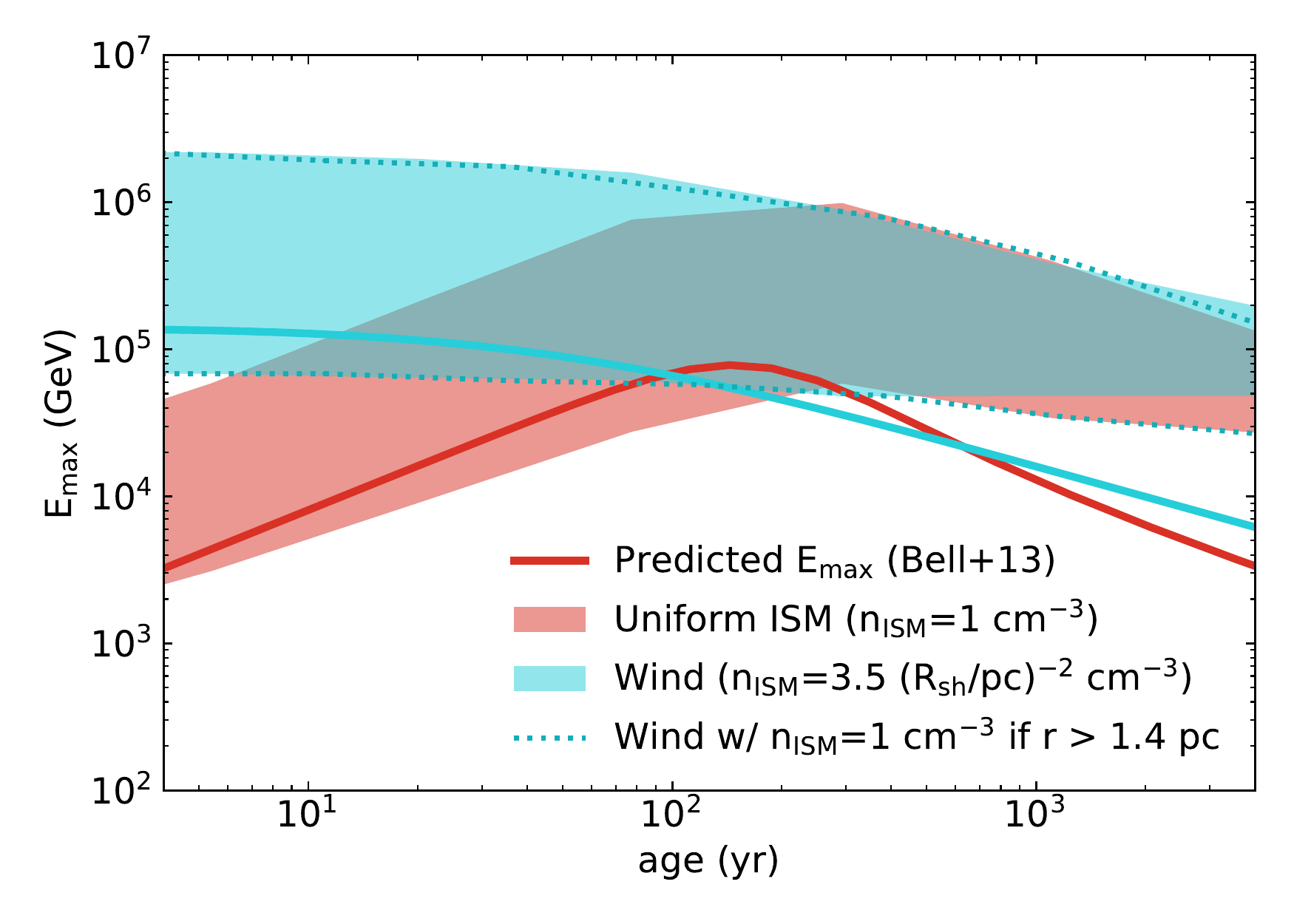}
    \caption{The maximum energy, $\Emax$, accelerated by our benchmark SNR as a function of time, expanding into a uniform medium (red band) and wind profile (blue band), as in Figure \ref{fig:Spectra}. The widths of the bands correspond to uncertainties in the nature of CR-driven magnetic field amplification (i.e., the lower (upper) limit assumes diffusing (escaping) particles drive amplification). Blue dotted lines indicate how our wind prediction would change if that wind had a finite size of 1.4 pc, corresponding to a shock age of $\sim 200$ yr}. Solid lines give the single-zone $\Emax$ prediction put forth in \citep{bell+13}. While this prediction yields good agreement with our results at early times, we predict a higher $\Emax$ at late times due to the presence of old populations of particles accelerated when the SNR was expanding more rapidly.
    \label{fig:EmaxTime}
\end{figure}

A more detailed picture of the temporal evolution of $\Emax$ can be found in Figure \ref{fig:EmaxTime}, which plots $\Emax$ as a function of shock age. 
Bands span the range between our limiting assumptions regarding the nature of magnetic field amplification, with red (blue) corresponding to expansion into the uniform (wind) profiles used in Figure \ref{fig:Spectra}. As Figure \ref{fig:EmaxTime} shows, the ambient medium strongly impacts the temporal evolution of $\Emax$, both because the amplified magnetic field depends on $\nism$ and, more importantly, because the density profile regulates the evolution of the shock radius and velocity. 

More specifically, recalling Equation \ref{eqn:Emax} for the instantaneous $\Emax$ along with the fact that, in a uniform medium, $R_{\rm sh} \propto t$ during the ejecta-dominated phase and $R_{\rm sh} \propto t^{2/5}$ during the Sedov-Taylor phase, we approximate,
\begin{equation}
\label{eqn:Emaxoft_uni}
\Emax \propto 
\begin{cases} 
      t & t \leq t_{\rm ST}  \\
      t^{-3/5} & t > t_{\rm ST}.
\end{cases}
\end{equation}
Here, $t_{\rm ST}$ denotes the onset of the Sedov-Taylor phase. For this reason, in a uniform medium, $\Emax$ reaches a peak around $t = t_{\rm ST}$.

Alternatively, in a wind profile, $R_{\rm sh} \propto t^{2/3}$ during the Sedov-Taylor phase.
Also accounting for the dependence of $\Emax$ on $\nism$ (which in turn goes as $R_{\rm sh}^{-2}$), we approximate,
\begin{equation}
\label{eqn:Emaxoft_wind}
\Emax \propto 
\begin{cases} 
      \text{constant} & t \leq t_{\rm ST}  \\
      t^{-5/6} & t > t_{\rm ST}. 
\end{cases}
\end{equation}
In other words, in a wind profile, $\Emax$ can only increase at very early times when it is still limited by the finite age of the system.
Note that this period of increasing $\Emax$ is much shorter than the evolutionary timescale of an SNR, meaning that, for the vast majority of its life, an SNR expanding into a wind profile exhibits a constant or decreasing $\Emax$.
However, due to very high densities at small radii, the highest value of $\Emax$ accelerated by SNRs expanding into wind profiles tends to exceed that accelerated by SNRs expanding into uniform media.

Of course, the ambient media surrounding real SNRs, particularly those of core-collapse SNe, can be more complicated than the simple profiles considered in this work. Detailed models of such environments, as well as the corresponding SNR evolutions and emissions, have been considered in the literature \cite[e.g., ][]{das+22, sushch+22, kobashi+22}. As the aim of this paper is to provide a more general estimate of the relationship between $\Emax$ and observationally inferrable shock parameters, we will not consider such complex scenarios. Rather, we encourage the reader to refer to Figure \ref{fig:EmaxVsh} and Equation \ref{eqn:Emax_empirical} for predictions of $\Emax$ that, while affected by the density of the ambient medium, are not highly sensitive to it.

That being said, to confirm that a more complex SNR environment would not produce results in conflict with Figure \ref{fig:EmaxVsh} or Equation \ref{eqn:Emax_empirical}, we test a modified version of the wind profile shown in Figures \ref{fig:Spectra} and \ref{fig:EmaxTime}, in which the wind has a finite size of $r_{\rm w} = 1.4$ pc \citep[corresponding to a wind mass of $\sim 1 M_{\odot}$, see, e.g.,][]{weaver+77}. Beyond 1.4 pc, the shock evolves in a uniform ISM of density 1 cm$^{-3}$. This scenario, shown as dotted lines in Figure \ref{fig:EmaxTime}, leads to significant deceleration of the shock at the transition to the uniform ISM (which occurs when the shock age is roughly 200 yr). However, as Figure \ref{fig:EmaxTime} shows, in terms of $\Emax$, introducing a finite wind size simply means that, outside of $r_{\rm w}$, $\Emax$ behaves as it would in the uniform case, with a relatively smooth transition owing to multi-zone effects (namely, the contributions of particle populations accelerated at earlier times).

We also compare our $\Emax$ results to the calculation presented in \cite{bell+13}, which sets $\Emax$ by requiring that the current of escaping particles of energy $E = \Emax$ be sufficient to produce strong magnetic field amplification via the non-resonant streaming instability. 
This approximation gives an instantaneous maximum energy that scales similarly to the relations described in the preceding paragraphs (see their Equation 6), and is displayed as solid lines in Figure \ref{fig:EmaxTime}.

Note that while our results are in good agreement with those of \cite{bell+13} at early times, they diverge after $t \sim t_{\rm ST}$. This divergence also occurs when comparing our results to the scaling relations shown in Equations \ref{eqn:Emaxoft_uni} and \ref{eqn:Emaxoft_wind}, and arises from the fact that both \cite{bell+13} and Equations \ref{eqn:Emaxoft_uni} and \ref{eqn:Emaxoft_wind} consider only a single population of particles. 
However, in our multi-zone framework, we account for the presence of old populations of particles with an $\Emax$ that may differ from the maximum energy currently being accelerated by the shock. 
As Equations \ref{eqn:Emaxoft_uni} and \ref{eqn:Emaxoft_wind} show, for $t > t_{\rm ST}$, these old populations have a larger $\Emax$ than the current instantaneous one, leading to a slowed decline of the cumulative maximum energy.
In other words, by considering the overall evolution of the shock, we find that older shocks may produce $\gamma$-ray signatures that point toward higher $\Emax$ than would be predicted from a single-zone model.

Finally, we present $\Emax$ as a function of a more readily observable parameter, $\vsh$, in Figure \ref{fig:EmaxVsh}, for a variety of modeled SNRs. As in Figure \ref{fig:Spectra}, the left (right) column corresponds to the case in which diffusing (escaping) particles drive magnetic field amplification, and the top (bottom) row corresponds to expansion in to a uniform (wind) profile. However, we also consider a wider range of shock parameters, as introduced in Section \ref{subsec:hydro}: $\nism \in [10^{-2}, 10^2]$ cm$^{-3}$ in the uniform case and $\nism \in [10^{-1}, 10]\times 3.5(R/\text{pc})^{-2}$ cm$^{-3}$ in the wind case. We also consider SNRs with $E_{\rm SN} = 10^{52}$ erg and/or $M_{\rm ej} = 5 M_{\odot}$ (wind case only).

As expected from Equations \ref{eqn:Emaxoft_uni} and \ref{eqn:Emaxoft_wind}, faster shocks have higher $\Emax$, as do shocks expanding into denser media, though this density dependence is mild (note that the normalization of $\nism$ is given by the color scale in Figure \ref{fig:EmaxVsh}). Thus, for shocks that have reached the Sedov-Taylor stage, $\vsh$---which is more easily inferred from observations than age---serves as a good proxy for $\Emax$. However, this relationship breaks down for young, ejecta-dominated SNRs, which have roughly constant $\vsh$ and thus an $\Emax$ that is set by $R_{\rm sh}$. This is especially true in the uniform case, in which $\Emax$ rises prior to $t_{\rm ST}$. To emphasize this issue, points in Figure \ref{fig:EmaxVsh} corresponding to $t < t_{\rm ST}$ are outlined in black.

Thus, the results shown in Figure \ref{fig:EmaxVsh} are best summarized with an empirical relationship that includes $\vsh$, $R_{\rm sh}$, and, to a lesser extent, $\nism$. 
We approximate this relationship as,
\begin{equation}
\label{eqn:Emax_empirical}
    \Emax \simeq \alpha \bigg(\frac{\nism}{\text{cm}^{-3}}\bigg)^\frac{1}{2}\bigg(\frac{\vsh}{10^3 \text{ km s}^{-1}}\bigg)^2\bigg(\frac{R_{\rm sh}}{\text{pc}}\bigg)\text{ GeV}
\end{equation}
where  $\alpha = 7\times 10^2$ if diffusing particles drive magnetic field amplification and $\alpha = 1.5\times 10^4$ if escaping particles drive magnetic field amplification. 
Note that this expression retains the same scalings with $R_{\rm sh}$ and $\nism$ as Equation \ref{eqn:Emax}, but has a slightly weaker dependence on $\vsh$ ($\Emax \propto \vsh^2$ instead of $\vsh^{5/2}$). 
This weakened velocity dependence approximates the multi-zone effects described previously, namely, that old populations of particles can contribute to the cumulative $\Emax$.

\section{Discussion} \label{sec:discussion}

We will now discuss our results in the context of particle acceleration up to PeV energies, i.e., the approximate energy of the CR ``knee."

\subsection{SNRs as PeVatrons}

As demonstrated in the preceding section, SNRs can only accelerate PeV particles under select circumstances:
\begin{enumerate}
    \item Escaping particles must drive magnetic field amplification.
    \item The shock must be expanding relatively quickly ($\vsh \gtrsim 10^4$ km s$^{-1}$ for $\nism = 1$ cm$^{-3}$).
    \item In the absence of a fast shock ($\vsh \gtrsim 10^4$ km s$^{-1}$), the ambient number density must be $\gg 1$ cm$^{-3}$.
\end{enumerate}
These conditions are comparable to those presented in the literature \citep[e.g.,][]{murase+11, bell+13, cardillo+15, marcowith+18, cristofari+20, cristofari+21}, which also find that only a small subset of fast SNRs expanding into dense media can be PeVatrons. Such stringent requirements may explain the dearth of observed PeVatrons that can be definitively associated with SNRs \citep[e.g.,][]{cao+21}. However, in contrast to recent works in the literature that completely rule out SNRs as PeVatrons \cite[e.g.,][]{brose+22} we find that it is possible for SNRs to at least contribute to---though perhaps not saturate---the CR knee.

That being said, the fact remains that typical historical SNRs are likely incapable of accelerating PeV particles, meaning that other astrophysical accelerators may contribute to the CR spectrum at the knee. Promising candidates include microquasars \citep[e.g.,][]{abeysekara+18}, star clusters \citep[e.g.,][]{aharonian+19, bykov+20}, and superbubbles \citep[e.g.,][]{parizot+04}.
In these pictures, shocks are still likely responsible for particle acceleration, meaning that the formalism and scaling relations presented in this work may be applicable. 
However, in many cases, the termination (i.e., reverse) shock is invoked as the primary acceleration site, necessitating a modified prescription for particle escape.

\subsection{Considerations for PeVatron Searches}

A notable result from our work is the fact that SNRs and other astrophysical shocks may exhibit $\gamma$-ray and neutrino signatures of PeV particles after they are no longer accelerating them. In this case, one might expect to see $\sim 100$ TeV $\gamma$-rays from SNRs approaching $t \simeq 100$ yr. Of course, the question remains whether such PeV particles would remain confined and, if not, how far they would propagate away from their accelerator.

After acceleration, a particle will remain confined within an SNR provided that, downstream of the shock, diffusion is insufficient to overcome advection  \citep[see, e.g.,][]{drury10}. 
In other words, we require that the advection timescale, $\tau_{\rm adv} \simeq R_{\rm sh}/u_2$ (where $u_2 \simeq \vsh/4$ is the velocity of the downstream plasma in the frame of the shock), be less than the diffusion timescale, $\tau_{\rm diff} \simeq R_{\rm sh}^2/D(E)$, i.e., we require $D(E) < u_2 R_{\rm sh}$. 
This requirement holds for all particles except those very close to $E = \Emax$, since $D(\Emax) \sim R_{\rm sh} \vsh$. For shocks expanding into uniform media, $R_{\rm sh} \vsh \propto t^{-1/5}$ after $t_{\rm ST}$, meaning that confined particles will eventually escape, but only after the shock has evolved for a long time.
For shocks expanding into wind profiles, $R_{\rm sh} \vsh \propto t^{1/3}$ after $t_{\rm ST}$, meaning that initially-confined particles are likely to remain so. 
In short, a shock that accelerates PeV particles may be able to confine them even after it is no longer capable of accelerating them.

Of course, $D(E)$ may also grow with time if, for example, magnetic turbulence is damped. However, even in the case that PeV particles quickly escape their accelerators, they may produce detectable $\gamma$-ray signatures in the vicinity. 
Namely, taking the canonical diffusion coefficient in our Galaxy, $D(E) \simeq 3\times 10^{28}(E/\text{GeV})^{1/3}$ cm$^2$ s$^{-1}$ \citep[i.e., the maximum possible value of $D(E)$, see, e.g.,][]{maurin+14}, PeV particles will only diffuse $\sim 100$ pc after $\sim 1000$ yr. 
This radius becomes even smaller in the case of suppressed diffusion near CR sources \citep[e.g.,][]{fujita+10, fujita+11}.
Moreover, in a uniform medium, the $\gamma$-ray and neutrino luminosities due to proton-proton collisions between these PeV particles and the ISM will remain constant as the CRs diffuse away, albeit smaller than it would have been had particles remained confined to the denser medium downstream of the shock. 
The exact luminosity of these $\gamma$-ray (and neutrino) ``halos" \citep[e.g.,][]{brose+21} will depend on the fraction of particles that escape, along with the nature of the nearby medium \citep[e.g., the presence of molecular clouds,][]{aharonian13}.
This consideration is especially important for water-Cherenkov observatories such as LHASSO and HAWC (and potentially the next generation of neutrino detectors, e.g., KM3NET, IceCube--Gen2, TRIDENT, P-One), which have comparatively lower resolutions than atmospheric Cherenkov telescopes and thus higher sensitivities to extended sources.

\begin{figure}[t]
    \centering
    \includegraphics[width = 0.47\textwidth, clip=true, trim = {5, 5, 5, 10}]{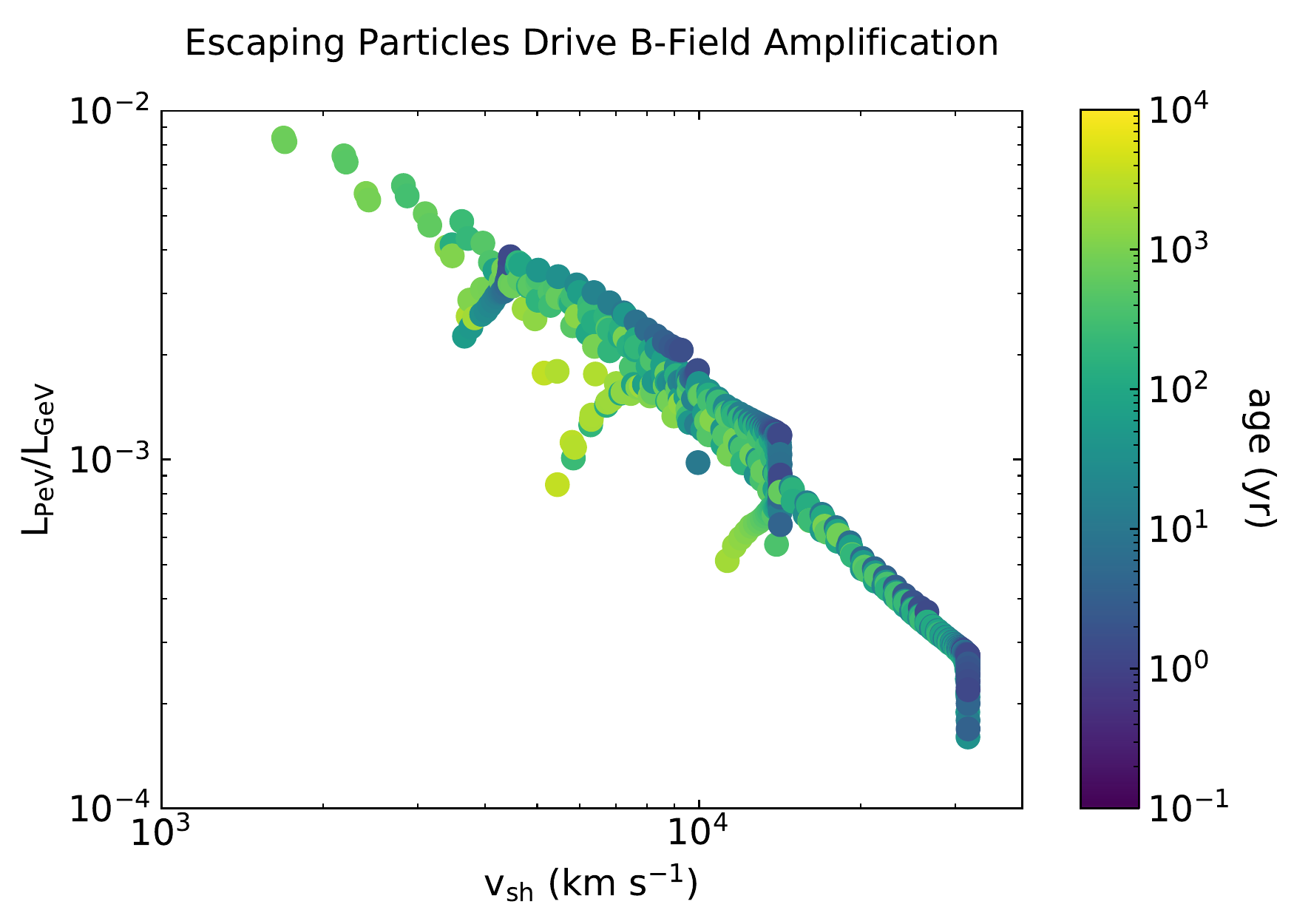}
    \caption{PeV to GeV proton luminosity ratio ($L_{\rm PeV}/L_{\rm GeV}$) as a function of $\vsh$ for SNR PeVatron candidates (i.e., data points from Figure \ref{fig:EmaxVsh} with $\Emax > 10^6$ GeV). Shock age is denoted with the color scale. Due to the \emph{postcursor} effect \citep[][see text for details]{diesing+21}, faster shocks---which are the best candidates to accelerate PeV particles---likely produce steeper spectra, and thus exhibit lower $L_{\rm PeV}/L_{\rm GeV}$. This effect is important for PeVatron searches that select observational targets based on their $\gamma$-ray luminosities at lower energies.}
    \label{fig:PeVRatio}
\end{figure}

Finally, it is worth noting that the \emph{postcursor} effect, introduced in Section \ref{subsec:Acceleration}, has implications for PeVatron searches with $\gamma$-ray observatories or neutrino detectors, particularly those that select sources based on their $\gamma$-ray luminosities at lower energies \citep[e.g.,][]{CTA23}. 
Namely, in this paradigm, faster shocks, which exhibit stronger amplified magnetic fields via the non-resonant instability, are expected to have faster-drifting magnetic fluctuations and thus steeper spectra. 
This expectation is also consistent with observations of both historical SNRs and young, extragalactic SNe \citep{diesing+21}.

Thus, if fast shocks tend to accelerate CRs with steeper spectra, the ostensible best candidates to be PeVatrons will have small CR luminosities at PeV energies ($L_{\rm PeV}$) relative to their CR luminosities at, say, GeV energies ($L_{\rm GeV}$). 
And, since hadronic $\gamma$-ray and neutrino emissions have the same spectral slope as that of the parent protons, this effect will be reproduced in observations.
We summarize this effect in Figure \ref{fig:PeVRatio}, which shows $L_{\rm PeV}/L_{\rm GeV}$ as a function of $\vsh$ for all modeled SNRs with $\Emax > 1$ PeV. 
Note that, regardless of the slope of the underlying particle population(s), $\gamma$-ray emission at TeV energies is likely to be hadronic in origin, since the maximum energy of leptons is limited by synchrotron losses \citep{corso+23}.

\section{Conclusion} \label{sec:conclusion}

In summary, we placed constraints on the maximum proton energy, $\Emax$, accelerated by an arbitrary astrophysical shock using a self-consistent, multi-zone model of particle acceleration, including the dynamical back-reaction of CRs on the shock as well as a prescription for magnetic field amplification that brackets theoretical uncertainties.
We presented our results in terms of parameters that can be constrained observationally, in particular the shock velocity (see Equation \ref{eqn:Emax_empirical}).
We also analyzed our results in the context of CR acceleration to PeV energies and discussed considerations for observational PeVatron searches.

Consistent with results presented in the literature, we find that typical historical SNRs cannot accelerate particles to PeV energies.
However, young, fast SNRs expanding into dense media can be PeVatrons if escaping particles drive magnetic field amplification.

We also find that old populations of particles can contribute to the cumulative spectrum accelerated by an astrophysical shock. 
This implies that SNRs and other cosmic accelerators may exhibit a higher $\Emax$ than they are currently capable of accelerating or, equivalently, that former PeVatrons may still produce $\gtrsim 100$ TeV $\gamma$-ray emission.

Finally, we note that, due to the drift of magnetic fluctuations with the local Alfv\'en speed downstream of astrophysical shocks (the \emph{postcursor}), fast shocks---which are the best candidates to produce PeV particles---have steeper spectra than their slower counterparts.
In other words, compared to a slower counterpart, a fast shock will have a smaller PeV luminosity relative to its luminosity at lower energies.
This consideration is important for observational PeVatron searches that select targets based on their GeV or TeV $\gamma$-ray luminosities.

These results serve as a reference to modelers seeking to estimate $\Emax$ for an arbitrary astrophysical shock, particularly in anticipation of the next generation of $\gamma$-ray and neutrino telescopes (e.g., The Cherenkov Telescope Array, IceCube--Gen2), which will better constrain the source(s) of PeV CRs in the coming years \citep[see, e.g.,][]{CTA23, sudoh+23}.

\acknowledgements
The author would like to thank Damiano Caprioli, Angela V. Olinto, Raffaella Margutti, Irina Zhuravleva, and Fausto Cattaneo for their valuable feedback. This research was partially supported by a William Rainey Harper Dissertation Fellowship and NSF grant AST-1909778.

\bibliographystyle{aasjournal}

\end{document}